\documentclass[conference,a4paper]{APSIPA2023}
\usepackage{multirow}
\usepackage{graphicx}
\usepackage{amsmath}
\usepackage[psamsfonts]{amssymb}
\usepackage{amsxtra}
\usepackage{threeparttable}

\usepackage{cite}
\usepackage{tabularx}

\usepackage{geometry}
\usepackage{url}
\usepackage{fancyhdr}

\fancypagestyle{firststyle} {
    \fancyhf{}
    \fancyhead[C]{2023 Asia Pacific Signal and Information Processing Association Annual Summit and Conference (APSIPA ASC)}

}

\begin{document}
\pagenumbering{gobble}
\title{Classification of Infant Sleep/Wake States: Cross-Attention among Large Scale Pretrained Transformer Networks using Audio, ECG, and IMU Data}

\author{%
\authorblockN{%
Kai Chieh Chang\authorrefmark{1}\authorrefmark{2},
Mark Hasegawa-Johnson\authorrefmark{1}\authorrefmark{2},
Nancy L. McElwain\authorrefmark{2}\authorrefmark{4},
Bashima Islam\authorrefmark{3}
}
\authorblockA{%
\authorrefmark{1}
Department of Electrical and Computer Engineering, University of Illinois at Urbana Champaign, USA}

\authorblockA{%
\authorrefmark{2}
Beckman Institute for Advanced Science and Technology, University of Illinois at Urbana-Champaign, USA  }

\authorblockA{%
\authorrefmark{4}
Department of Human Development and Family Studies, University of Illinois at Urbana-Champaign, USA  }

\authorblockA{%
\authorrefmark{3}
Department of Electrical and Computer Engineering, Worcester Polytechnic Institute, USA \\
E-mail: kcchang3@illinois.edu, jhasegaw@illinois.edu, mcelwn@illinois.edu, bislam@wpi.edu}

}

\maketitle
\thispagestyle{firststyle}
\pagestyle{fancy}

\begin{abstract}

  Infant sleep is critical to brain and behavioral development. Prior studies on infant sleep/wake classification have been largely limited to reliance on expensive and burdensome polysomnography (PSG) tests in the laboratory or wearable devices that collect single-modality data. To facilitate data collection and accuracy of detection, we aimed to advance this field of study by using a multi-modal wearable device, LittleBeats (LB), to collect audio, electrocardiogram (ECG), and inertial measurement unit (IMU) data among a cohort of 28 infants. We employed a 3-branch (audio/ECG/IMU) large scale transformer-based neural network (NN) to demonstrate the potential of such multi-modal data. We pretrained each branch independently with its respective modality, then finetuned the model by fusing the pretrained transformer layers with cross-attention. We show that multi-modal data significantly improves sleep/wake classification (accuracy = 0.880), compared with use of a single modality (accuracy = 0.732). Our approach to multi-modal mid-level fusion may be adaptable to a diverse range of architectures and tasks, expanding future directions of infant behavioral research. 
\end{abstract}

\section{Introduction}

Sleep is a physiological and behavioral process central to brain development during infancy \cite{nancy5, nancy6, nancy7, nancy8}. During the first six months of life, infants spend more time sleeping than awake \cite{nancy9}. Further, indices of infant sleep quality and quantity are associated with subsequent cognitive and language development \cite{nancy17, nancy18, nancy19, nancy20, nancy21, nancy22}, attention regulation \cite{nancy26, nancy27, nancy28}, social-emotional functioning \cite{nancy21, nancy26}, and physical health \cite{nancy33, nancy34}. Given the importance of infant sleep to development, automated and unobtrusive monitoring of sleep is critical. Thus, we focus on the task of sleep/wake classification in this study.

\begin{figure}[t]
    \begin{center}
    \includegraphics[width=3in]{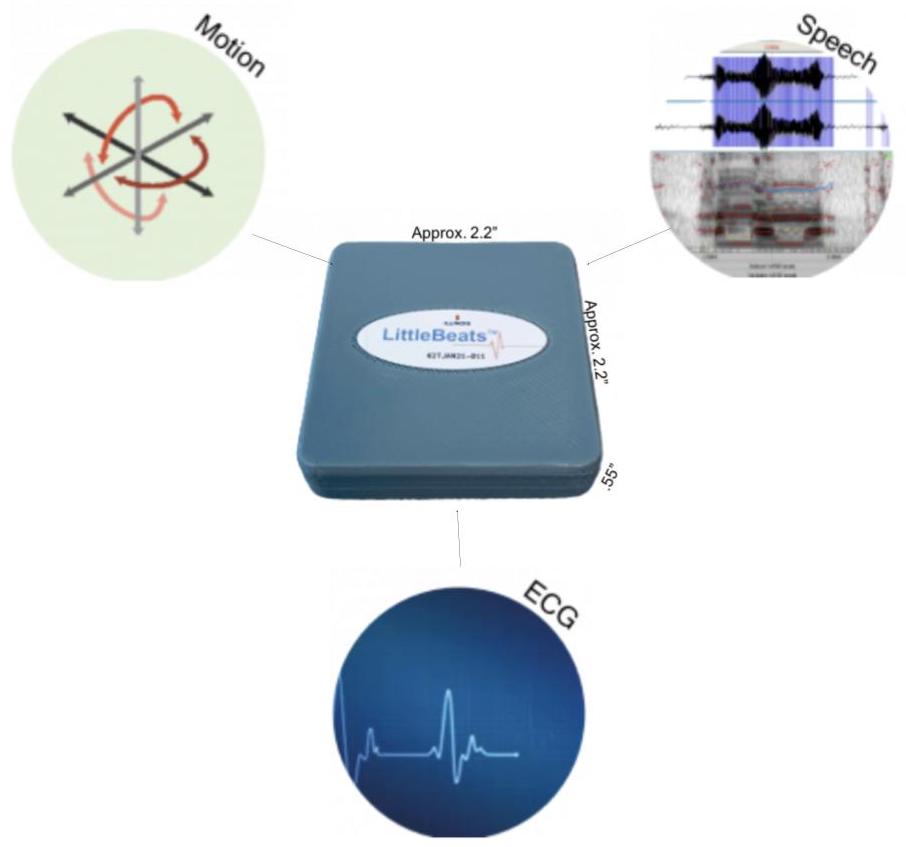}
    \caption{LittleBeats (center) is capable of collecting audio, ECG, and IMU data. Reproduced with permission from \cite{LB_Website}}
    \label{Fig:LittleBeatsSetup}
    \end{center}
\end{figure}

Laboratory-based polysomnography (PSG) is the gold-standard sleep assessment for adults and infants alike \cite{nancy45, nancy46}. However, given the high level of burden and expense of laboratory PSG, researchers have used wearable devices for sleep monitoring \cite{sleep_assessment_methods}. Notably, past studies of sleep/wake classification using wearable devices typically extract sleep features from single-modality data. For example, with actigraphy data, \cite{actigraphy_benchmark} carried out extensive studies on adult sleep/wake classification using traditional signal processing methods, classic machine learning methods, and several deep learning methods. Electrocardiogram (ECG) signals are also often used because stable QRS complex and longer R peak period can indicate sleep stages \cite{sleep_affect_ecg}. Studies such as \cite{sleepwake_ecg_cnn} used ECG data to classify adult sleep/wake with a 5-layer convolutional neural network (CNN) concatenated with 2 fully connected (FC) layers. A signal-processing-based approach was suggested by \cite{sleepwake_audio}, where audio recordings of adults' respiratory sounds were used to estimate sleep/wake likelihood from signal level features such as autocorrelation. All published studies of automatic sleep/wake classification use different datasets, therefore their accuracies are not strictly comparable, but the accuracies reported in all four of these studies are between 81\%  and 84\% (see Table \ref{table1} for details). A few studies have used bimodal data to further improve classification performance. Ref. \cite{sleepwake_respiratory+ecg} concatenated the frequency bins from respiratory signals and ECG to feed into a FC based network, achieving classification accuracy of 85.3\%. Ref. \cite{walch} uses random forest to classify sleep/wake from acceleration and heart rate data collected by Apple Watch, achieving 87.3\% accuracy. Recent work tend to focus on a more involved sleep classification with 5 stages. Ref. \cite{sleepwake_audiovideo} extracted infant vocalizations from audio recordings and physical motions from video recordings, and classified sleep stages using a random forest, resulting in an average accuracy near 85\%. Supratak et al.'s DeepSleepNet \cite{deepsleepnet} and their subsequent stage of the art (SOTA) TinySleepNet \cite{tinysleepnet} used only electroencephalogram (EEG) to reach 87.5\% in accuracy.

To achieve further improvements in accuracy, we leveraged trimodal data collected by an infant wearable device, LittleBeats (LB), as shown in Fig. \ref{Fig:LittleBeatsSetup}. LB can synchronously record infant sounds (via microphone [audio]), physiology (via 3-lead ECG), and motion (via inertial measurement unit [IMU]) for extended periods of time (8-10 hours) in the home context.

To perform sleep/wake classification using LB multi-modal data, we developed an ensemble of three large scale transformer networks pretrained on unimodal audio, ECG, and IMU datasets, and fine-tuned using trimodal data. This system has the potential to combine all important feature extraction done by the research mentioned above, including the extraction of vocalization and breathing patterns from audio recording, QRS complex features from ECG, and position and motion from IMU. To the best of our knowledge, our study is the first to access these three modalities jointly on sleep/wake classification. We show superior classification performance using all three modalities compared with any single modality or pairs of modalities combined.

In contributing to neural network (NN) architectures, we explore the benefits of pretraining large scale transformer networks on unlabeled audio, ECG, and IMU data, then finetuning a cross-attention based fusion architecture on a small LB labeled dataset. Unsupervised pretraining involves training a NN on a large, diverse dataset, which enables it to learn generic latent features that can be applied to a wide range of tasks. There is a wide range of pretraining schemes for audio, ECG, and IMU data. Specifically, wav2vec 2.0 (W2V2) \cite{wav2vec2} showed the benefits of pretraining audio using contrastive loss, achieving state of the art performance in tasks such as automatic speech recognition. Ref. \cite{wav2vec2_ecg} combined W2V2 and contrastive multi-segment coding to pretrain on ECG signals, achieving outstanding performance on cardiac arrhythmia classification and patient identification. LIMU-BERT \cite{limu-bert} is derived from bidirectional encoder representations from transformers (BERT) \cite{bert}, and is used to pretrain transformers with unlabeled IMU data. Inspired by the above architectures, we pretrained three branches of large scale transformer networks, one for each modality, and were able to extract latent features useful for sleep/wake classification.

To utilize features from all three modalities, we implemented a fusion strategy that relies on cross-attention among pretrained transformer layers. Multimodal fusion is often done by early fusion, where the input from each modality is concatenated and fed into the network as a single input, or late fusion, where each modality is processed separately and the outputs are combined at the final stage. We define early fusion as fusing right after feature extraction and not fusion of raw data. Much work has been done on these 2 fusion techniques. Ref. \cite{ecg_imu_early_fusion} employed early fusion by cascading learned features from ECG and IMU directly, and passed them through a dense network. Ref. \cite{early_late_fusion} described a simple late fusion of different branches by summing or averaging the logit outputs. Recently, intermediate fusion has become more common. This type of feature fusion happens in the middle of the NN, and uses a variety of architectures depending on the tasks. Ref. \cite{einv2} extensively used so-called ``cross-stitch" modules for soft parameter fusion, taking multiple linear weighted sums of each feature extraction layer as input for the next layer. Cross-attention based fusion techniques were explored in recent papers such as \cite{cross_attn_1} \cite{cross_attn_2}, where the attention layers take concatenated features from different modalities as input, and \cite{cross_attn_3}, where a single cross-attention layer is used to share information between branches. Our approach is innovative, first, in that it relies on the pretrained transformer layers from each branch, rather than training a feature sharing mechanism from scratch. Second, we fuse the three transformer networks by alternating self and cross-attention at different layers, which both preserves each branch's transformer features and incorporates attention from other modalities. 

In sum, in addition to being the first study to combine audio, ECG, and IMU for classification of infant sleep/wake states, we develop an innovative cross-attention based fusion for large scale pretrained transformer networks to combine three modalities. We not only reiterate the well known benefit of pretraining, but also demonstrate the ability of pretrained unimodal transformers to be fine-tuned using cross-attention-based multi-modal fusion to improve accuracy. We believe our work lays the groundwork for improved accuracy in a wide variety of signal processing tasks by the use of multimodal wearable devices such as LB.

\section{Method}

\begin{figure}[!t]
    \centering
    \includegraphics[width=3.5in]{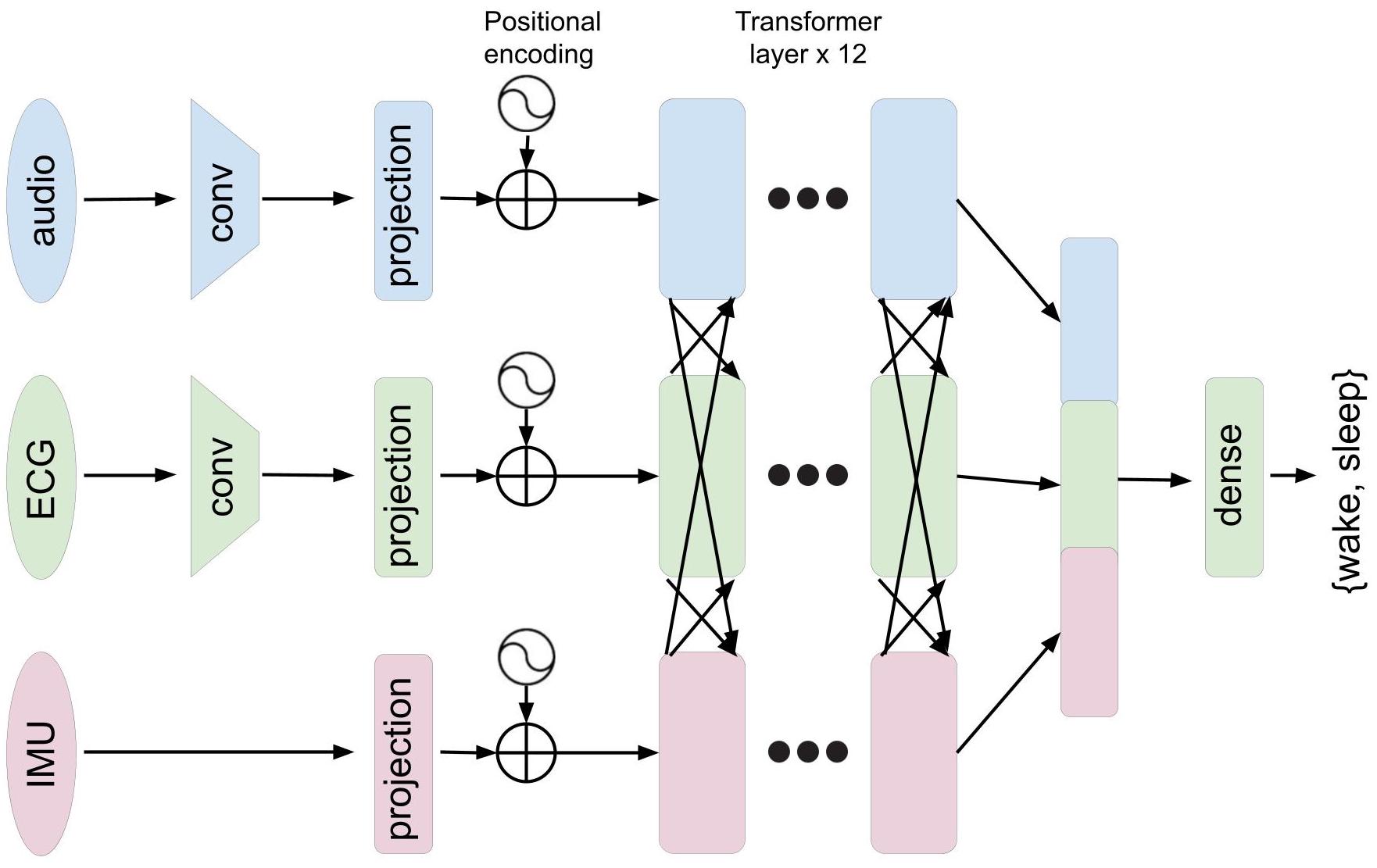}
    \caption{The LittleBeats Model Architecture. Each of the audio (top), ECG (center), and IMU (bottom) branches consists of feature extraction (convolutional and/or feedforward projection), positional encoding, and 12 transformer layers. Each branch is pretrained individually on unlabeled data. During labeled finetuning, the three branches are combined via cross-attention at the feature level. Their outputs are then concatenated and fed into dense layers to output logits.}
    \label{Fig:LB_model}
\end{figure}


Our architecture is an ensemble of three similarly structured transformer networks, with each branch targeting audio, ECG, or IMU data respectively, as shown in Fig. \ref{Fig:LB_model}. 

For the audio branch, we utilize a standard 12-layer W2V2 \cite{wav2vec2} network because W2V2 repeatedly shows superb performance in various speech-related tasks such as automatic speech recognition and phoneme recognition. A recent study \cite{jialu} showed an oracle W2V2 pretrained on 4300 hours of unlabeled infant-family audio data collected by LB or the Language ENvironment Analysis (LENA) system \cite{lena} increased performance of speech diarization and vocalization type classifications. As infant vocalization features may well include cues of sleep, we adopt pretrained weights from \cite{jialu} for the audio W2V2 branch.

As for ECG data, we were inspired by the similarity between speech and ECG. As early as 1940, researchers \cite{vocoder} were actively using impulse trains to imitate glottal excitation for speech generation. R-peaks in ECG data have a similar time domain structure, so a speech-centric NN should be able to learn ECG features. Speech and ECG differ in one respect: speech contains periodicity at many frequencies (pitch and formants), while ECG is dominated by the inter-beat interval, therefore ECG features may be learnable using less pre-training than speech. We pretrain a standard 12-layer W2V2 using 574 hours of unlabeled ECG data to utilize our observations above and to match the structure of the audio branch.

IMU data does not share similar structure as audio or ECG: an IMU signal is composed of parallel 3-axis accelerometer and 3-axis gyroscope signals, whose correlations are important for signal interpretation.  To take full advantage of the information contained in our IMU data, we used a pretraining paradigm based on LIMU-BERT \cite{limu-bert}.  In place of W2V2's CNN feature extraction layers, LIMU-BERT begins with a single fully-connected feature projection layer per input sample, followed by several layers of transformers. This network is pretrained with Masked Language Model objectives and a Span Masking mechanism. We pretrained 574 hours of IMU data collected from LB, on a modified LIMU-BERT architecture where we lifted the restriction that all transformers share weights, extended the number of layers to 12, and matched the transformer implementation to the one found in W2V2. 

\begin{figure}[!t]
    \centering
    \includegraphics[width=3.5in]{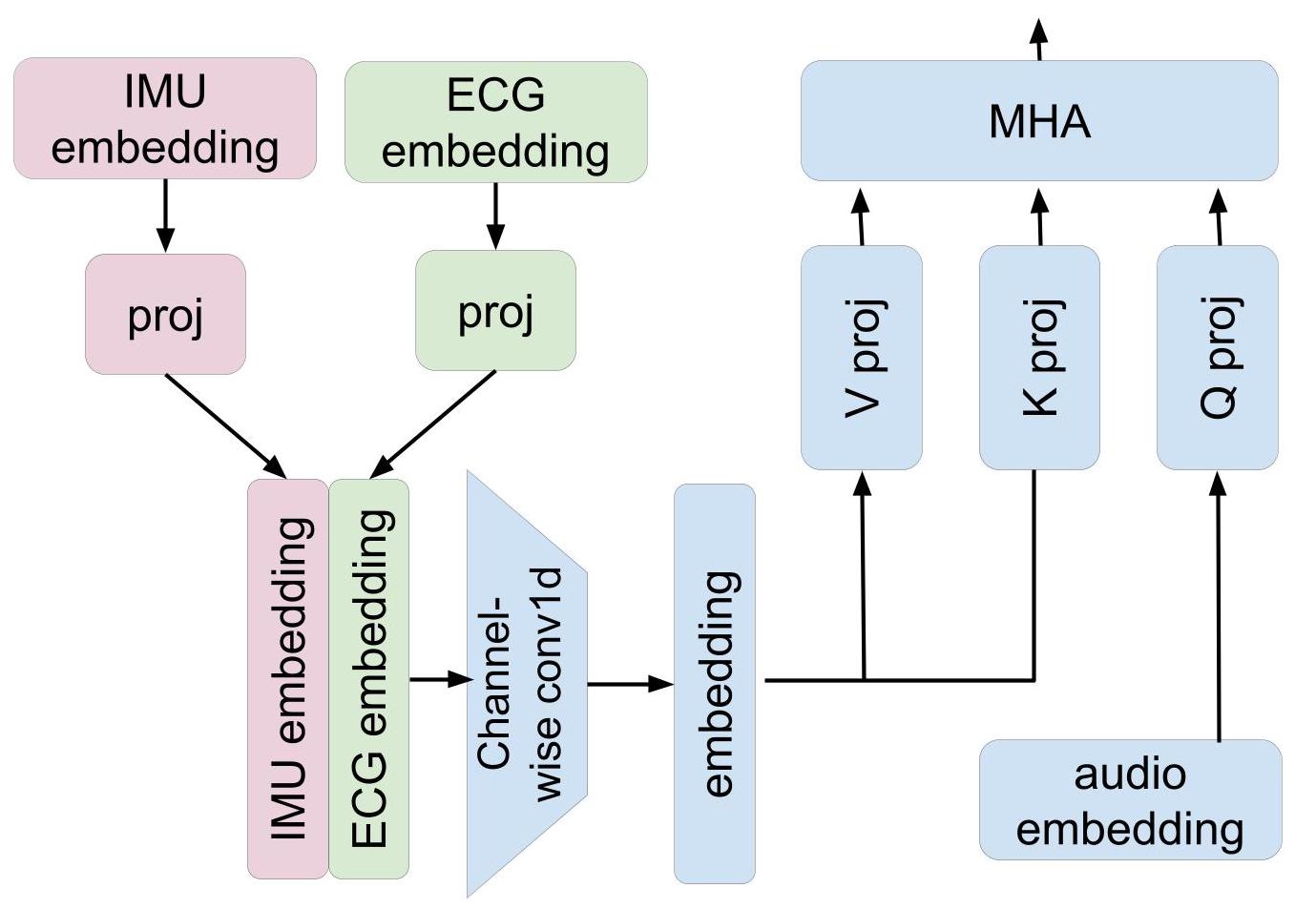}
    \caption{Cross-attention for the audio branch. Multiheaded attention layer (MHA) takes the audio embedding as query. IMU and ECG outputs from the other two branches first get linearly projected and concatenated into a 2-channel embedding. The embedding is reduced to 1-channel by a 1-d convolution layer and passed into MHA as key and value.}
    \label{Fig:cross_attention}
\end{figure}

To maximize the model's ability to learn across modalities, or to extend the transformer attention beyond a single modality, we utilized linear transformation and cross-attention every 4 transformer layers, as inspired by \cite{chang2021endtoend} and their work using cross-attention across audio channels. An example of the cross-attention mechanism for the audio branch is shown in Fig. \ref{Fig:cross_attention}. A cross-attention layer for the audio branch consists of multiple cross-attention heads, each of which takes its query from the previous layer's audio embedding, and its values and keys from a fusion of the previous layer's ECG and IMU branches.  ECG and IMU embeddings are first passed through two different linear layer projections producing an output with the frame rate and embedding dimension of the audio, then concatenated to create a 2 channel embedding. A channel-wise 1d convolutional layer is then used to create the vector that serves as key and value of the audio cross-attention layer.  The multiple audio cross-attention heads are then concatenated, added to a residual connection (from the previous audio layer), normalized, and passed through a feedforward layer.  The ECG and IMU cross-attention layers have the same architecture, with each modality receiving query inputs from its own previous layer, while each modality's key and value inputs come from the other two modalities. 

Mathematically, we can denote each hidden input embedding to the transformer layer as $\mathbf{H}_k \in \mathbb{R}^{N\times L_e}$, where the indices $(k,l,m)$ are any permutation of $\{\text{audio}, \text{ECG}, \text{IMU}\}$, $N$ is the number of samples in the input, and $L_e$ is the architecture embedding size. Let's define the query, key, and value projection of each branch to be $\mathbf{W}^{\{Q, K, V\}} \in \mathbb{R}^{L_e \times L_e}$. Let's constrain the discussion of attention regarding one head of one branch where $k=\text{audio}$. The attention layers can be self attention or cross-attention depending on the layer index $i$. We found using cross-attention at layers 1, 5, and 9 (i.e. $i\%4=1$ for layers $0\le i\le 11$) gives the best performance.


\begin{align*}
    \text{SelfAttn}_{i,k} = 
    \text{softmax}\left(
   \frac{\mathbf{H}_{k}\mathbf{W}_i^Q  \left(\mathbf{H}_{k}\mathbf{W}_i^K\right)^T}{\sqrt{L_e}}\right) \mathbf{H}_{k}\mathbf{W}_i^V
\end{align*}

\begin{align*}
    \text{CrossAttn}_{i,k} = 
    \text{softmax}\left(
   \frac{\mathbf{H}_{k}\mathbf{W}_i^Q  
   \left(\mathbf{H}_{\text{cross}}
    \mathbf{W}_i^K\right)^T}{\sqrt{L_e}}\right)
       \mathbf{H}_{\text{cross}}
    \mathbf{W}_i^V
\end{align*}

\begin{align*}
       \mathbf{H}_{\text{cross}}
= \mathbf{A} \left[ 
    \begin{matrix} 
    \text{proj}(\mathbf{H}_{l}^T)\\
    \text{proj}(\mathbf{H}_{m}^T)
    \end{matrix} \right]
\end{align*}
where $\mathbf{A} \in \mathbb{R}^{L_e \times 2}$ is a trainable mixing weight. This discussion is applicable to the other heads and other branches. 

The three branches' outputs are concatenated and passed through three dense layers with ReLU activations to produce binary logits for sleep and wake. With the pretrained weights loaded onto each branch and the ensemble in place, we use frame-wise cross entropy loss to finetune the network.

\section{Experiment}
\subsection{Dataset}
Participants were recruited via study flyers distributed to local community organizations as well as online listservs serving families of young children. In the context of larger studies of child behavioral development, study coordinators provided participating families with a LB or LENA kit, along with instructions for conducting associate recordings in the home. Families were asked to complete 2 to 3 daylong (8+ hours) recordings over a 2-week period. All study procedures were approved by the Institutional Review Board at the University of Illinois Urbana-Champaign. 

Unlabeled data included (a) 4300hr of audio home recordings (1000hr collected by LB and 3200hr by LENA) from 245 families with children under 5 years of age (see \cite{jialu} for further details), as well as (b) 574hr of ECG data and (c) 574hr of IMU data collected by LB from 28 families with children (54\% female) under 5 years of age (Mean = 26 months, Range: 3-65 months). For the labeled data, we gathered 68.5hr of synchronized audio, ECG, and IMU data from the same 28 families. Four trained human annotators labeled infant sleep and wake states from 1.5hr segments of LB audio files. Annotation of sleep states was facilitated by referring to (a) parental reports of their child’s activities while wearing the LB device in the home, (b) visual inspection of the wav form of the audio file, and (c) listening to audio recordings for indicators of infant sleep (e.g., slowed steady breathing, no vocalizations or movement). 

We randomly divide the labeled data into training set (50hr, 25 families), validation set (4.5hr, 1 family), and testing set (14hr, 2 families). Data from families/infants in the training, validation, and testing datasets did not overlap.

\begin{figure}[!t]
    \centering
    \includegraphics[width=3.5in]{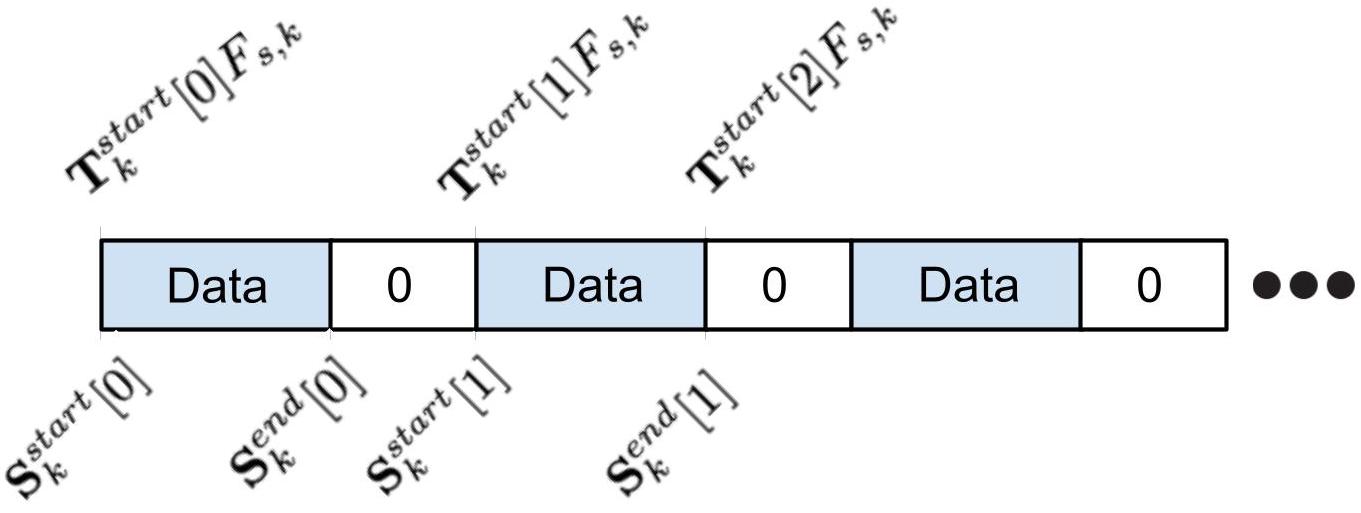}
    \caption{Zeroing data according to timestamp. For each data segment, the recorded UTC time multiplied by sampling frequency ($F_{s, k} * (\mathbf{T}_{k}^{\text{end}}[i] - \mathbf{T}_{k}^{\text{start}}[i])$) is longer than the recorded number of samples ($\mathbf{S}_{k}^{\text{end}}[i] - \mathbf{S}_{k}^{\text{start}}[i]$). The missing data are simply filled with zeros.}
    \label{Fig:data_zeroing}
\end{figure}

\subsection{Data Preprocessing}
As a first step, we synchronized data across modalities. LB records each modality in slightly different time segments. For each segment, LB records the timestamps in absolute UTC start time and end time, and the associated start and end sample index. Formally, let's define 
\begin{align*}
    \mathbf{T}_{k}^{\text{start}}, \mathbf{T}_{k}^{\text{end}}, \mathbf{S}_{k}^{\text{start}}, \mathbf{S}_{k}^{\text{end}} \in \mathbb{R}^{L_{k}}
\end{align*}
where for each modality, each element in $\mathbf{T}$ is the absolute UTC start or end time, each element in $\mathbf{S}$ is starting or ending data sample index, $L$ is the number of data chunks.

LB is designed to function continuously despite variability in the number of nonzero IMU samples, therefore it sometimes fails to record all samples in a given audio or ECG segment.  Such missed samples can be detected by comparing the known ground truth duration of a segment, $F_{s, k} * (\mathbf{T}_{k}^{\text{end}}[i] - \mathbf{T}_{k}^{\text{start}}[i])$, to the number of samples it contains, $(\mathbf{S}_{k}^{\text{end}}[i] - \mathbf{S}_{k}^{\text{start}}[i])$, where $F_{s, k}$ is the sampling frequency for modality $k$. When an incomplete segment is detected, we fill in the missing data with zeros. We obtain $\mathbf{\hat{Z}}_{k}$ where each of the vectors has length $F_{s, k} * (\mathbf{T}_{k}^{\text{end}}[i] - \mathbf{T}_{k}^{\text{start}}[i])$. This is shown in Fig. \ref{Fig:data_zeroing}.

Timestamps of different modalities are approximately but not precisely aligned, because the writing to SD card on LB is asynchronous.  In order to make multimodal processing possible,  we simply discard the data that did not overlap according to the timestamp and truncate  $\mathbf{\hat{Z}}_{k}$ to $\mathbf{Z}_{k} \in \mathbb{R}^{L}$ where L is the same across all modalities. $\mathbf{Z}_{k}$ is then segmented into 30s segments $\mathbf{X}_{k}$. 

Human annotators mark the beginning and ending of each period of sleep.  These times are used to assign, to each 30s segment $\mathbf{X}_k$, a binary label $\mathbf{Y}_{k} \in \{0, 1\}$ where 0 is wake and 1 is sleep.  If both sleep and wake labels are present in segment $k$, $\mathbf{Y}_{k}$ is set equal to the label with longer duration.  

\subsection{Metric}
Because this is a binary classification task, we evaluated standard accuracy, precision, recall, F1-score, and Cohen's kappa $\kappa$.

\begin{table*}[!t]
\renewcommand{\arraystretch}{1.3}
\caption{Results Comparing Architectures with Different Number and Types of Modalities.  Results in the first section are our work; results in the second section are baselines computed using the same test set; results in the third section are copied from the corresponding reference papers, and were therefore computed using incomparable test sets.}
\label{table1}
\centering
\begin{tabular}{c c c c c c c c}
\hline
Test Set & Architecture & Modality & $\uparrow$ Accuracy & $\uparrow$ Precision & $\uparrow$ Recall & $\uparrow \text{F}1$ & $\uparrow$ Kappa\\
\hline
LB 68.5 hr & W2V2-mod & Audio & 0.797 & 0.794 & 0.909 & 0.848 & 0.546\\
 & W2V2-mod & ECG & 0.732 & 0.704 & \textbf{0.980} & 0.819 & 0.347\\
 & LIMU-BERT-mod & IMU & 0.786 & 0.837 & 0.815 & 0.826 & 0.549\\
 & Proposed & Audio + ECG + IMU & \textbf{0.880} & \textbf{0.881} & 0.934 & \textbf{0.907} & \textbf{0.741}\\
\hline
LB 68.5 hr & FFT + Dense \cite{sleepwake_respiratory+ecg} mod & Audio + ECG & 0.444 & 0.729 & 0.168 & 0.273 & 0.0523\\
 & CNN \cite{sleepwake_ecg_cnn} mod & ECG & 0.565 & 0.494 & 0.434 & 0.634 & 0.0991\\
 & Random Forest \cite{walch} & ECG + IMU & 0.717 & 0.729 & 0.799 & 0.762 & 0.410\\
 & CNN + LSTM \cite{tinysleepnet} & ECG & 0.843 & 0.825 & 0.778 & 0.748 & 0.639\\
\hline
Adult 7.1 hr & Signal Processing \cite{sleepwake_audio} & Audio  & 0.833 & - & 0.922 & - & 0.508\\
Adult 37k hr & LSTM \cite{actigraphy_benchmark} & Actigraphy  & 0.831 & 0.816 & 0.914 & 0.855 & -\\
Adult 292 hr & FFT + Dense \cite{sleepwake_respiratory+ecg} & Respiratory + ECG & 0.853 & - & - & - & -\\
Adult 110 hr & CNN \cite{sleepwake_ecg_cnn} & ECG & 0.815 & 0.478 & 0.930 & 0.360 & 0.260\\
Adult 244 hr & Random Forest \cite{walch} & ECG + IMU & 0.873 & - & 0.895 & - & 0.396\\

\hline
\end{tabular}
\end{table*}

\subsection{Implementation Details}

Audio sample rate is downsampled from 24000Hz to 16000Hz and ECG sample rate is upsampled from 2381Hz to 16000Hz, to match the default sampling rate of W2V2. IMU data are kept at a sampling rate of 150Hz because upsampling to 16000Hz might introduce too many artifacts. We did not experiment with data interpolation to fill out the zeros during time synchronization nor with data augmentation to diversify the data.

For audio and ECG W2V2 branch, we used almost the same baseline structure as described in \cite{wav2vec2}, with 12 transformer layers, hidden size 768, intermediete size 3072 and 16 attention heads. Convolution feature extraction structure remain the same with hidden size 512, kernel size [10,3,3,3,3,2,2] and strides [5,2,2,2,2,2,2]. For IMU branch, the feature extraction is done by a linear layer, a normalization layer, and adding a position embedding. The decoding part has 12 transformer layers, hidden size 72, intermediate size 144 and 4 attention heads. Outputs of the three branches are concatenated into a vector of size 1608, passed through a dense structure as follows: a linear layer from size 1608 to size 1608, ReLU, linear layer from size 1608 to 804, dropout of rate 0.1, linear layer from 804 to 2. 

Each experiment was run for 2 epochs with batch size 16, with the standard Adam optimizer. This was done on a single RTX A6000 GPU. Training code was based on Huggingface Transformers and the code is available for inspection and further development.~\footnote{This footnote will be replaced by a github URL containing an open-source implementation upon acceptance of the manuscript.}

\subsection{Baseline Implementation} \label{Baseline Implementation}
The majority of sleep/wake classification baselines did not release code, so we implemented some of their algorithms with modification based on their description in the paper. Specifically, for \cite{sleepwake_respiratory+ecg} we replaced the respiratory input with LB audio data, and trained with gradient descent instead of Levenberg–Marquardt back-propagation algorithm. For \cite{sleepwake_ecg_cnn}, we used increasing strides ([1, 2, 3, 4, 5, 6]) for the CNN layers instead of using [2, 2, 2, 2, 2, 2] because each LB ECG segments has more samples than theirs. We also modified the dropout value to 0.1 based on empirical testing. \cite{walch} used 3-axis acceleration and heart rate data collected on Apple Watch as their dataset. We calculated heart rate at each recorded segment from LB ECG data, using the python package HeartPy. We then input our timestamp, acceleration data, and calculated heart rate to their released code. Finally, state of the art sleep stage classifier \cite{tinysleepnet} released their code online. We used ECG as input to the model instead of EEG in the original paper.

\section{Results}

\subsection{Benefits of Three Modalities}

The first section in Table \ref{table1} shows how using all modalities together compares to using each of the modalities alone for the purpose of sleep/wake classification. The single modality architecture takes each branch in Fig. \ref{Fig:LB_model}, with pretrained weights, and adds the final three FC layers for classification. The proposed architecture is the entire model pretrained on unlabeled data and finetuned with proposed cross-attention at every 4 layers. We see that, by quite a large margin, using all three modalities together as proposed, the network is able to learn more about infants' sleep patterns and achieve better performance across the board than using just a single modality. Note that while recall for the ECG branch outperforms the proposed network by 5\%, its precision is lower by 20\%, suggesting that the unimodal ECG network is a worse classifier since it too frequently classifies segments as ``sleep.''

The second section in Table \ref{table1} presents results of key baseline implementation on the LB dataset, as described in Section \ref{Baseline Implementation}. Because \cite{sleepwake_respiratory+ecg} and \cite{sleepwake_ecg_cnn} have not released their code and \cite{walch} and \cite{tinysleepnet} are not designed to work strictly on ECG data, results need to be interpreted with caution. However, we can still observe that our proposed model using all three modalities outperforms all key baselines, which only use one or two modalities. 

Numbers in the last section in Table \ref{table1} are results reported in other papers, using other datasets, and are therefore not strictly comparable to the numbers in the first two sections.  The classifiers summarized by these results are sleep/wake classifiers using audio only \cite{sleepwake_audio}, ECG only \cite{sleepwake_ecg_cnn}, ECG and respiratory signals \cite{sleepwake_respiratory+ecg},  motion \cite{actigraphy_benchmark}, and ECG and IMU \cite{walch}. While all datasets are different and comparison of the accuracies is therefore not theoretically justifiable, such a comparison nevertheless supports the conclusion displayed in the top half of the table, viz., that sleep/wake classification performed using three modalities is more accurate than sleep/wake classification performed using only one or two modalities. 


\subsection{Benefits of Pretraining}

Table \ref{table2} shows how pretraining affects the performance of the proposed architecture. Here we configured the model to use all three branches, but only load the pretrained weights for specific modalities. Cross-attention is applied at every 
$4^{\text{th}}$ layer
for the 12 layers of transformers. We see that every time we add pretrained weights for a particular branch, we have a performance increase across all metrics. This reinforces our understanding of pretrained networks. By pretraining each branch using the unlabeled data, then fine-tuning the entire system together against a small number of labeled data, we are able to get better performance.

\subsection{Fusion Techniques}
\begin{table}[!t]
\renewcommand{\arraystretch}{1.3}
\caption{Results when Pretraining Different Branches of the Architecture}
\label{table2}
\centering
\begin{tabularx}{\linewidth}{c c c c c c}
\hline
Pretrained & $\uparrow$ Accuracy & $\uparrow$ Precision & $\uparrow$ Recall & $\uparrow \text{F}1$ & $\uparrow$ Kappa\\
\hline
None & 0.797 & 0.829 & 0.850 & 0.839 & 0.565\\
IMU & 0.837 & 0.874 & 0.863 & 0.868 & 0.655\\
Audio+IMU & 0.871 & 0.858 & \textbf{0.949} & 0.901 & 0.715\\
ECG+IMU & 0.875 & 0.866 & 0.944 & 0.904 & 0.726\\
All & \textbf{0.880} & \textbf{0.881} & 0.934 & \textbf{0.907} & \textbf{0.741}\\

\hline
\end{tabularx}
\end{table}

\begin{table}[!t]
\renewcommand{\arraystretch}{1.3}
\caption{Results of Early, Late, and the Proposed Fusion Technique}
\label{tablefusion}
\centering
\begin{tabular}{c c c c c c}
\hline
Fusion & $\uparrow$ Accuracy & $\uparrow$ Precision & $\uparrow$ Recall & $\uparrow \text{F}1$ & $\uparrow$ Kappa\\
\hline
Early & 0.779 & 0.857 & 0.773 & 0.813 & 0.544\\
Late & 0.798 & 0.781 & \textbf{0.947} & 0.856 & 0.530\\
Proposed & \textbf{0.880} & \textbf{0.881} & 0.934 & \textbf{0.907} & \textbf{0.741}\\

\hline
\end{tabular}
\end{table}

Our discussion is incomplete without comparison to early and late fusion techniques discussed in the literature \cite{early_late_fusion}. Table \ref{tablefusion} presents performance of 2 fusion methods against the proposed cross-attention-based intermediate fusion technique. Early fusion is inspired by \cite{ecg_imu_early_fusion}, where convolutional features extracted from ECG and IMU data are concatenated with anthropometric data as input to a downstream neural network. In our variation, we concatenate the three outputs from the feature extractor for each branch as shown in Fig. \ref{Fig:LB_model}, skipping the transformer layers, and pass through the dense network for fine-tuning. As for late fusion, we leave in the transformer layers, triplicate the FC layers to generate logits for each modality, and average the three logits for evaluation. The proposed cross-attention fusion achieved better performance in almost every metric except that it has a lower recall than late fusion. Similar to the trend found in Table \ref{table1}, this high recall for late fusion results from a disproportional confusion matrix, as shown by late fusion's low precision score. Therefore, the proposed fusion architecture is still favorable in general.

\subsection{Ablation Study}

Table \ref{table3} shows the effect of using cross-attention at different layers. All models were fine-tuned on labeled multimodal data after loading pretrained weights for all three modalities separately. ``-" means the architecture did not converge. We empirically found that cross-modal attention once every 4 layers ($i\%4=1$) gives the best performance across the board. Many models with cross-modal attention fail to converge: convergence failure occurs with cross-modal attention every second layer, and with cross-modal attention in the middle four layers ($4\le i<8$) or the last four layers ($8\le i<12$).  Cross-modal attention in the first four layers ($0\le i<4$) converged to a system with accuracy lower than that of the system with no cross-modal attention.  The sensitivity of cross-modal attention to architectural configuration was an unexpected result.  We defer this to future studies.

\begin{table}[!t]
\renewcommand{\arraystretch}{1.3}
\caption{Result with Cross-Attention Applied at Various Layers.  The equation in the first column specifies the layers in which cross-modal attention is applied, e.g., $i\%4=1$ means that cross-modal attention was performed in layers 1, 5, and 9, where layers are numbered $0\le i\le 11$.}
\label{table3}
\centering
\begin{tabularx}{\linewidth}{c c c c c c}
\hline
Cross & $\uparrow$ Accuracy & $\uparrow$ Precision & $\uparrow$ Recall & $\uparrow F1$ & $\uparrow$ Kappa\\
\hline
None & 0.765 & 0.788 & 0.851 & 0.818 & 0.487\\
i\%2=1 & - & - & - & - & -\\
i\%4=1 & \textbf{0.880} & \textbf{0.881} & \textbf{0.934} & \textbf{0.907} & \textbf{0.741}\\
i\%6=1 & 0.836 & 0.870 & 0.865 & 0.868 & 0.652\\
$0 \leq i < 4$ & 0.755 & 0.811 & 0.791 & 0.801 & 0.484\\
$4 \leq i < 8$ & - & - & - & - & -\\
$8 \leq i < 12$ & - & - & - & - & -\\

\hline
\end{tabularx}
\end{table}

\section{Conclusion}
With the development of multi-modal wearable devices, we are able to gather synchronized audio, ECG, and IMU data for the task of infant sleep/wake classification. We demonstrated the best classification performance when using all three modalities compared with our own single modality network or single/double modality network found in the literature. In addition, we developed an ensemble of large scale pretrained transformer neural network, by fusing the pretrained transformer layers with cross-attention at every 4 layers.  This fusion method is not limited to the task of wake/sleep classification, but seems likely to generalize successfully to any multi-modal network in which all modalities have the same number of pretrained transformer layers. Our work presents exciting directions for multi-modal studies of infant and child development.

\section*{Acknowledgment}

We would like to thank the families who participated in this research, as well as Jordan Bodway and Jenny Baldwin for their assistance with data collection and processing. This work was supported by funding from the National Institute of Mental Health (R21MH112578), National Institute of Drug Abuse (R34DA050256), the National Institute of Food and Agriculture (ILLU-793-368), and the Personalized Nutrition Initiative and Center for Social and Behavioral Science at the University of Illinois at Urbana-Champaign through the Seed Grant program. 

\bibliographystyle{IEEEtran}
\bibliography{ref.bib}

\begin{thebibliography}{10}
\providecommand{\url}[1]{#1}
\csname url@samestyle\endcsname
\providecommand{\newblock}{\relax}
\providecommand{\bibinfo}[2]{#2}
\providecommand{\BIBentrySTDinterwordspacing}{\spaceskip=0pt\relax}
\providecommand{\BIBentryALTinterwordstretchfactor}{4}
\providecommand{\BIBentryALTinterwordspacing}{\spaceskip=\fontdimen2\font plus
\BIBentryALTinterwordstretchfactor\fontdimen3\font minus
  \fontdimen4\font\relax}
\providecommand{\BIBforeignlanguage}[2]{{%
\expandafter\ifx\csname l@#1\endcsname\relax
\typeout{** WARNING: IEEEtran.bst: No hyphenation pattern has been}%
\typeout{** loaded for the language `#1'. Using the pattern for}%
\typeout{** the default language instead.}%
\else
\language=\csname l@#1\endcsname
\fi
#2}}
\providecommand{\BIBdecl}{\relax}
\BIBdecl

\bibitem{nancy5}
A.~R. Tarullo, P.~D. Balsam, and W.~P. Fifer, ``Sleep and infant learning,''
  \emph{Infant Child Dev.}, vol.~20, no.~1, pp. 35--46, Jan. 2011.

\bibitem{nancy6}
M.~S. Blumberg, A.~J. Gall, and W.~D. Todd, ``The development of sleep-wake
  rhythms and the search for elemental circuits in the infant brain,''
  \emph{Behav. Neurosci.}, vol. 128, no.~3, pp. 250--263, Jun. 2014.

\bibitem{nancy7}
R.~E. Dahl, ``The regulation of sleep and arousal: Development and
  psychopathology,'' \emph{Dev Psychopathol}, vol.~8, no.~3, pp. 3--27, 1996.

\bibitem{nancy8}
E.~Bathory and S.~Tomopoulos, ``Sleep regulation, physiology and development,
  sleep duration and patterns, and sleep hygiene in infants, toddlers, and
  preschool-age children,'' \emph{Curr. Probl. Pediatr. Adolesc. Health Care},
  vol.~47, no.~2, pp. 29--42, Feb. 2017.

\bibitem{nancy9}
B.~C. Galland, B.~J. Taylor, D.~E. Elder, and P.~Herbison, ``Normal sleep
  patterns in infants and children: a systematic review of observational
  studies,'' \emph{Sleep medicine reviews}, vol.~16, no.~3, p. 213—222, June
  2012.

\bibitem{nancy17}
A.~Hupbach, R.~L. Gomez, R.~R. Bootzin, and L.~Nadel, ``Nap-dependent learning
  in infants,'' \emph{Dev. Sci.}, vol.~12, no.~6, pp. 1007--1012, Nov. 2009.

\bibitem{nancy18}
K.~Horv{\'a}th, K.~Myers, R.~Foster, and K.~Plunkett, ``Napping facilitates
  word learning in early lexical development,'' \emph{J. Sleep Res.}, vol.~24,
  no.~5, pp. 503--509, Oct. 2015.

\bibitem{nancy19}
E.~Dearing, K.~McCartney, N.~L. Marshall, and R.~M. Warner, ``Parental reports
  of children's sleep and wakefulness: longitudinal associations with cognitive
  and language outcomes,'' \emph{Infant Behav. Dev.}, vol.~24, no.~2, pp.
  151--170, Feb. 2001.

\bibitem{nancy20}
A.~Bernier, S.~M. Carlson, S.~Bordeleau, and J.~Carrier, ``Relations between
  physiological and cognitive regulatory systems: infant sleep regulation and
  subsequent executive functioning,'' \emph{Child Dev.}, vol.~81, no.~6, pp.
  1739--1752, Nov. 2010.

\bibitem{nancy21}
M.~Ednick, A.~P. Cohen, G.~L. McPhail, D.~Beebe, N.~Simakajornboon, and R.~S.
  Amin, ``A review of the effects of sleep during the first year of life on
  cognitive, psychomotor, and temperament development,'' \emph{Sleep}, vol.~32,
  no.~11, pp. 1449--1458, Nov. 2009.

\bibitem{nancy22}
A.~Scher, ``Infant sleep at 10 months of age as a window to cognitive
  development,'' \emph{Early Hum. Dev.}, vol.~81, no.~3, pp. 289--292, Mar.
  2005.

\bibitem{nancy26}
A.~Sadeh, G.~De~Marcas, Y.~Guri, A.~Berger, L.~Tikotzky, and Y.~Bar-Haim,
  ``Infant sleep predicts attention regulation and behavior problems at 3-4
  years of age,'' \emph{Dev. Neuropsychol.}, vol.~40, no.~3, pp. 122--137,
  2015.

\bibitem{nancy27}
M.~Thunstr{\"o}m, ``Severe sleep problems in infancy associated with subsequent
  development of attention-deficit/hyperactivity disorder at 5.5 years of
  age,'' \emph{Acta Paediatr.}, vol.~91, no.~5, pp. 584--592, 2002.

\bibitem{nancy28}
F.~V. O'Callaghan, A.~Al~Mamun, M.~O'Callaghan, A.~Clavarino, G.~M. Williams,
  W.~Bor, H.~Heussler, and J.~M. Najman, ``The link between sleep problems in
  infancy and early childhood and attention problems at 5 and 14 years:
  Evidence from a birth cohort study,'' \emph{Early Hum. Dev.}, vol.~86, no.~7,
  pp. 419--424, Jul. 2010.

\bibitem{nancy33}
J.-P. Chaput, C.~E. Gray, V.~J. Poitras, V.~Carson, R.~Gruber, C.~S. Birken,
  J.~E. MacLean, S.~Aubert, M.~Sampson, and M.~S. Tremblay, ``Systematic review
  of the relationships between sleep duration and health indicators in the
  early years (0-4 years),'' \emph{BMC Public Health}, vol.~17, no. Suppl 5, p.
  855, Nov. 2017.

\bibitem{nancy34}
L.~Matricciani, C.~Paquet, B.~Galland, M.~Short, and T.~Olds, ``Children's
  sleep and health: A meta-review,'' \emph{Sleep Med. Rev.}, vol.~46, pp.
  136--150, Aug. 2019.

\bibitem{LB_Website}
\BIBentryALTinterwordspacing
``About littlebeats.'' [Online]. Available:
  \url{https://littlebeats.hdfs.illinois.edu/about-littlebeats/}
\BIBentrySTDinterwordspacing

\bibitem{nancy45}
A.~Sadeh, ``Iii. sleep assessment methods,'' \emph{Monogr. Soc. Res. Child
  Dev.}, vol.~80, no.~1, pp. 33--48, Mar. 2015.

\bibitem{nancy46}
S.~E. Beck and C.~L. Marcus, ``Pediatric polysomnography,'' \emph{Sleep Med.
  Clin.}, vol.~4, no.~3, pp. 393--406, Sep. 2009.

\bibitem{sleep_assessment_methods}
A.~Sadeh, ``Iii. sleep assessment methods,'' \emph{Monogr. Soc. Res. Child
  Dev.}, vol.~80, no.~1, pp. 33--48, Mar. 2015.

\bibitem{actigraphy_benchmark}
J.~Palotti, R.~Mall, M.~Aupetit, M.~Rueschman, M.~Singh, A.~Sathyanarayana,
  S.~Taheri, and L.~Fernandez-Luque, ``Benchmark on a large cohort for
  sleep-wake classification with machine learning techniques,'' Jun 2019.

\bibitem{sleep_affect_ecg}
C.~Cajochen, J.~Pischke, D.~Aeschbach, and A.~A. Borbély, ``Heart rate
  dynamics during human sleep,'' \emph{Physiology \& Behavior}, vol.~55, no.~4,
  pp. 769--774, 1994.

\bibitem{sleepwake_ecg_cnn}
J.~Malik, Y.-L. Lo, and H.~tieng Wu, ``Sleep-wake classification via
  quantifying heart rate variability by convolutional neural network,''
  \emph{Physiological Measurement}, vol.~39, no.~8, p. 085004, aug 2018.

\bibitem{sleepwake_audio}
E.~Dafna, A.~Tarasiuk, and Y.~Zigel, ``\BIBforeignlanguage{en}{Sleep-wake
  evaluation from whole-night non-contact audio recordings of breathing
  sounds},'' p. e0117382, Feb 2015.

\bibitem{sleepwake_respiratory+ecg}
W.~Karlen, C.~Mattiussi, and D.~Floreano, ``Sleep and wake classification with
  ecg and respiratory effort signals,'' \emph{IEEE Transactions on Biomedical
  Circuits and Systems}, vol.~3, no.~2, pp. 71--78, 2009.

\bibitem{walch}
O.~Walch, Y.~Huang, D.~Forger, and C.~Goldstein, ``{Sleep stage prediction with
  raw acceleration and photoplethysmography heart rate data derived from a
  consumer wearable device},'' \emph{Sleep}, vol.~42, no.~12, 08 2019, zsz180.

\bibitem{sleepwake_audiovideo}
S.~Cabon, F.~Porée, A.~Simon, B.~Met-Montot, P.~Pladys, O.~Rosec, N.~Nardi,
  and G.~Carrault, ``Audio- and video-based estimation of the sleep stages of
  newborns in neonatal intensive care unit,'' \emph{Biomedical Signal
  Processing and Control}, vol.~52, pp. 362--370, 2019.

\bibitem{deepsleepnet}
A.~Supratak, H.~Dong, C.~Wu, and Y.~Guo, ``{DeepSleepNet}: A model for
  automatic sleep stage scoring based on raw single-channel {EEG},''
  \emph{{IEEE} Transactions on Neural Systems and Rehabilitation Engineering},
  vol.~25, no.~11, pp. 1998--2008, nov 2017.

\bibitem{tinysleepnet}
A.~Supratak and Y.~Guo, ``Tinysleepnet: An efficient deep learning model for
  sleep stage scoring based on raw single-channel eeg,'' \emph{2020 42nd Annual
  International Conference of the IEEE Engineering in Medicine \& Biology
  Society (EMBC)}, pp. 641--644, 2020.

\bibitem{wav2vec2}
A.~Baevski, H.~Zhou, A.~Mohamed, and M.~Auli, ``wav2vec 2.0: A framework for
  self-supervised learning of speech representations,'' 2020.

\bibitem{wav2vec2_ecg}
J.~Oh, H.~Chung, J.~myoung Kwon, D.~gyun Hong, and E.~Choi, ``Lead-agnostic
  self-supervised learning for local and global representations of
  electrocardiogram,'' 2022.

\bibitem{limu-bert}
H.~Xu, P.~Zhou, R.~Tan, M.~Li, and G.~Shen, ``Limu-bert: Unleashing the
  potential of unlabeled data for imu sensing applications,'' in
  \emph{Proceedings of the 19th ACM Conference on Embedded Networked Sensor
  Systems}, ser. SenSys '21.\hskip 1em plus 0.5em minus 0.4em\relax New York,
  NY, USA: Association for Computing Machinery, 2021, p. 220–233.

\bibitem{bert}
J.~Devlin, M.-W. Chang, K.~Lee, and K.~Toutanova, ``Bert: Pre-training of deep
  bidirectional transformers for language understanding,'' 2019.

\bibitem{ecg_imu_early_fusion}
Z.~Ni, T.~Wu, T.~Wang, F.~Sun, and Y.~Li, ``Deep multi-branch two-stage
  regression network for accurate energy expenditure estimation with ecg and
  imu data,'' p. 3224–3233, Oct 2022.

\bibitem{early_late_fusion}
K.~Gadzicki, R.~Khamsehashari, and C.~Zetzsche, ``Early vs late fusion in
  multimodal convolutional neural networks,'' in \emph{2020 IEEE 23rd
  International Conference on Information Fusion (FUSION)}, 2020, pp. 1--6.

\bibitem{einv2}
Y.~Cao, T.~Iqbal, Q.~Kong, F.~An, W.~Wang, and M.~D. Plumbley, ``An improved
  event-independent network for polyphonic sound event localization and
  detection,'' 2021.

\bibitem{cross_attn_1}
L.~Ying, H.~Yu, J.~Wang, Y.~Ji, and S.~Qian, ``Multi-level multi-modal
  cross-attention network for fake news detection,'' \emph{IEEE Access},
  vol.~9, pp. 132\,363--132\,373, 2021.

\bibitem{cross_attn_2}
X.~Wei, T.~Zhang, Y.~Li, Y.~Zhang, and F.~Wu, ``Multi-modality cross attention
  network for image and sentence matching,'' in \emph{2020 IEEE/CVF Conference
  on Computer Vision and Pattern Recognition (CVPR)}, 2020, pp.
  10\,938--10\,947.

\bibitem{cross_attn_3}
X.~Song, H.~Chao, X.~Xu, H.~Guo, S.~Xu, B.~Turkbey, B.~J. Wood, T.~Sanford,
  G.~Wang, and P.~Yan, ``Cross-modal attention for multi-modal image
  registration,'' \emph{Medical Image Analysis}, vol.~82, p. 102612, 2022.

\bibitem{jialu}
J.~Li, M.~Hasegawa-Johnson, and N.~L. McElwain, ``Towards robust family-infant
  audio analysis based on unsupervised pretraining of wav2vec 2.0 on
  large-scale unlabeled family audio,'' 2023.

\bibitem{lena}
D.~Xu, J.~A. Richards, and J.~Gilkerson, ``Automated analysis of child phonetic
  production using naturalistic recordings,'' \emph{J. Speech Lang. Hear.
  Res.}, vol.~57, no.~5, pp. 1638--1650, Oct. 2014.

\bibitem{vocoder}
H.~Dudley, ``The vocoder—electrical re-creation of speech,'' \emph{Journal of
  the Society of Motion Picture Engineers}, vol.~34, no.~3, pp. 272--278, 1940.

\bibitem{chang2021endtoend}
F.-J. Chang, M.~Radfar, A.~Mouchtaris, B.~King, and S.~Kunzmann, ``End-to-end
  multi-channel transformer for speech recognition,'' 2021.

\end{thebibliography}



\end{document}